# A Novel Approach to Probability

Oded Kafri

## Abstract

When $P$ indistinguishable balls are randomly distributed among $L$ distinguishable boxes, and considering the dense system $P \gg L$, our natural intuition tells us that the box with the average number of balls $P/L$ has the highest probability and that none of boxes are empty; however in reality, the probability of the empty box is always the highest. This fact is with contradistinction to sparse system $P \ll L$ (i.e. energy distribution in gas) in which the average value has the highest probability. Here we show that when we postulate the requirement that all possible configurations of balls in the boxes have equal probabilities, a realistic "long tail" distribution is obtained. This formalism when applied for sparse systems converges to distributions in which the average is preferred. We calculate some of the distributions resulted from this postulate and obtain most of the known distributions in nature, namely: Zipf's law, Benford's law, particles energy distributions, and more. Further generalization of this novel approach yields not only much better predictions for elections, polls, market share distribution among competing companies and so forth, but also a compelling probabilistic explanation for Planck's famous empirical finding that the energy of a photon is $h\nu$.

# Introduction

Probability theory is an important branch of mathematics because of its huge economic significance. Since our knowledge about the future is limited, the most effective decisions rely on statistics.

Usually it is assumed that similar underlying events should have a similar resultant expectation value. From the law of large numbers [1] we expect that a large number of trials will yield the average value. From the central limit theorem [2] we expect that the scattered deviations from average will be distributed as in a Gaussian curve (the normal distribution).

Accordingly, when pollsters try to predict the results of a poll, i.e. elections, they treat each party individually. They find its expected number of seats from a small sample. Then they combine the results of all the parties together and obtain the expected distribution of the seats among the parties in the election. The *a priori* assumption of any pole is that the distribution of the seats among the parties, if the people vote randomly, is equal.

Based on the second law of thermodynamics, we suggest an alternative assumption that should be made, namely, that any configuration of seats among the parties should have an equal probability. When we calculate the distribution of seats under this assumption, we obtain an unequal distribution of seats among parties which is very similar to that obtained in real elections, see Fig. 3 below. Nevertheless, each of the parties has an equal *a priori* probability to win each of the unequal number of seats of the distribution.

To generalize the analysis we call each party a box, and the available seats in parliament, balls. In statistical physics a box is called a state, and a possible configuration of the balls in the boxes is called a microstate. The logarithm of the number of microstates is called entropy[3]. The second law of thermodynamics states that a system in equilibrium has maximum entropy. Maximum entropy exists when all the microstates have an equal probability and all the states have an equal probability.

We distinguish between two kinds of statistical systems:

*Dense systems*, namely, systems in which the number of balls is bigger than the number of boxes.

*Sparse systems*, in which the number of balls is much smaller than the number of boxes and/or no more than one ball can exist in a box.

For example, elections are dense systems, since there are more seats than parties. Similarly, a best sellers' list is dense as the number of readers is greater than the number of the books. In wealth distribution, there are more dollar bills than people, there are more surfers per second than sites in the internet, and more listeners than songs, etc. It seems that many of the human's social activities can be described by dense systems statistics.

On the other hand lotteries are a sparse systems, since the number of winners is smaller than the number of tickets sold. The sparse systems statistics are very useful in describing distributions in space, because in many cases two objects cannot occupy the same space, i.e. two buildings cannot coexist in the same area. Similarly, two atoms or molecules cannot coexist in the same state.

In order to visualize the difference between the classic approach and the new approach to probability let us look at 12 balls distributed randomly in 3 distinct boxes. The average is 4, therefore most people will assume that the best guess is to find 4 balls in any randomly chosen box. The second best guesses are 3 and 5 balls in a box. Yet, in reality, if you have to gamble your best bet should be an empty box.

The guess of 4 balls is based on the law of large number and the guess of 3 and 5 balls are based on the central limit theorem. However, the probability to find 4 balls in a box is only 10%, 3 balls is 11% , 1 ball is 13%, and the most probable is an empty box which has 14% probability. These probabilities were calculated according to the new approach and it reflects the known fact that there is a higher probability to be poor than to be rich.

The reason for that counterintuitive result is found in the microstates of this system. For example: there are only 3 different configurations in which boxes with 12 balls exist; namely, 12,0,0; 0,12,0; and 0,0,12. For these configurations, there are only 3 full boxes with 12 balls' while the number of empty boxes is double that, namely 6. Empty boxes are found, in addition, in many other configurations as well. This is the intuitive explanation why the expectation of the number of the boxes having $n$ balls is monotonically increasing as $n$ decreases.

It is not easy to count the number of appearance of the boxes with $n$ balls in all the microstates, as it requires tedious combinatorial calculations because the number of microstates is exploding very fast. Moreover, the numbers are specific to any values of balls and boxes. However, the techniques of maximizing the entropy, which was first suggested by Planck yields a proven good approximations. The reason why entropy maximization yields the best solution to these problems is because the entropy is defined in equilibrium, in which the number of microstates is maximum because each microstate has an equal probability.

From the requirement of an equal probability to each microstate, we can derive the relative number of boxes $\varphi(n)$ that have $n$ balls. We call $\varphi(n)$ frequency, and the balls distribution function is calculated by $\rho(n) = n\varphi(n)$.

The relation between $n$ and $\varphi(n)$ is derived by maximizing the entropy, using the Lagrange multipliers technique.

We can extend the theory more by giving the balls a "property". This is done by letting the balls have energy, assets, links etc.; In this case the property distribution calculations are similar to the calculations of the energy distribution among particles in statistical mechanics.

These kinds of probabilities are of great importance in statistics, econometrics, information theory and physics. Here we address the calculation of these probabilities in a unified way. In statistical mechanics this kind of analysis is called maximum entropy [4], in information theory - Shannon limit [5], and in classical physics - thermodynamic equilibrium [3].

As is shown hereafter, this analysis yields for integer number of balls in a box: Zipf's law [6,7], Benford's law [8,9,10], Elections votes distribution, Planck's law [11], Bose-Einstein distribution [12,13], Fermi-Dirac distribution and Maxwell-Boltzmann Distributions [19,20,21].

When the balls have a property, a chemical potential usually appears in the distribution functions.

It should be noted that maximum entropy techniques were used in probability theory and in econometrics [14, 15]. However, the entropy's expression that was used in these publications namely, Boltzmann-Gibbs entropy [16] is approximated for infinitely small $P/L$, as opposed to the examples discussed in this paper in which $P \gg L$. The Boltzmann-Gibbs entropy is applicable to sparse systems and yields the Maxwell-Boltzmann distribution in which the boxes with average values of balls have the highest probabilities. For example, in ideal gas the highest number of molecules have the average energy. Similarly the highest number of people have average height. This is with contradistinction to the example of 12 balls distributed in 3 boxes in which the highest probability is to find an empty box, similarly to the wealth distribution in which is it easier to find a poor man than to find a rich man.

## Entropy, frequency and distribution for integer number of balls

The model consists of identical indistinct balls and identical distinct boxes. The balls have no interaction between themselves and are distributed randomly among the boxes in a way that maximizes the entropy of the system under various constraints. We show how various statistical laws and distributions are derived using this model.

*Dense Systems*

Here there is no inherent limitations to the number of the balls in a box. Namely, it can be any number up to the total number of balls or any limit that we artificially impose.

We consider two constraints: the first is the total number of balls $P$ in the boxes and the second is the total number of the boxes. When the balls have a property, the constraint on the number of balls is replaced by a constraint on the total amount of property.

When the balls are pure logical entities like numbers, we say that the balls have no chemical potential. A reader of a book or a seat in election has no chemical potential. When the balls consist of a measurable quantity, the balls have chemical potential. For example, balls of a given weight or a given amount of energy have chemical potential. Similarly, balls with assets, have chemical potential.

Suppose that we have $P$ identical indistinct balls and $L$ identical distinct boxes. The number of the distinguishable ways to arrange the balls in the boxes (microstates) is given by the number of the configurations as was first suggested by Planck,

$$W = \frac{(L+P-1)!}{(L-1)!P!} \tag{1}$$

We designate $n = P/L$ and apply the Stirling formula $\ln n! \cong n \ln n - n$, we obtain [11] an expression for the entropy $S$,

$$S = \ln W \cong L[(n+1)\ln(n+1) - n \ln n]. \tag{2}$$

We designate the relative number of the boxes with $n$ balls by $\varphi(n)$ and therefore the number the different groups of boxes having $n$ balls is $N = L/\varphi(n)$, or,

$$L = N\varphi(n) \text{ and accordingly the number of balls is } P = Nn\varphi(n) \tag{3}$$

Where $L$ and $P$ of Eq.'s (3) are the two constraints of the system. The left one is the constraint on the total number of boxes $L$, and the right one is the constraint on the total number of balls $P$.

<u>*Example 1*</u> **The distribution of balls in the boxes: Zipf's law and Planck-Benford law**

To write the Lagrange function we substitute the entropy from Eq. (2) and add the constraints for $P$ and $L$ from Eq. (3). The Lagrange equation $F(n)$ is,

$$F(n) \cong L[(n+1)\ln(n+1) - n \ln n] + \beta'[P - Nn\varphi(n)] + \alpha[L - N\varphi(n)] \tag{4}$$

Where $\beta'$ is a Lagrange multiplier of the constraint of the number of balls, and $\alpha$ is the Lagrange multiplier of the constraint of the number of boxes. The maximum entropy solution is obtained when

$\partial F(n)/\partial n = 0$.

Since $L\partial[(n+1)\ln(n+1) - n \ln n]/\partial n = L \ln \frac{n+1}{n}$ and $-\beta' N \partial[n\varphi(n)]\partial n = -\beta' N \varphi(n)$, and $\alpha N \partial[\varphi(n)]\partial n = 0$, we obtain from Eq. (4),

$$n = \frac{1}{\exp(\beta\varphi(n))-1} \tag{5}$$

Where we designate $\beta' \equiv \beta L/N$.

Eq. (5) is the probabilistic Planck equation. It should be noted here that $L$ is canceled out in the derivation and we remain with knowledge only about the boxes with balls. Where $\beta\varphi(n) \ll 1$, then $\exp(\beta\varphi(n)) \cong 1 + \beta\varphi(n)$ and we obtain,

$n\varphi(n) \cong \beta^{-1}$,

which is Zipf's Law. From Eq.(5) it is seen that Zipf's law is obtained when the number of balls in a box $n$ is large. Zipf's law is known in its outcome that,

$\frac{\varphi(1)}{\varphi(2)} = \frac{\varphi(2)}{\varphi(4)} = \frac{\varphi(4)}{\varphi(8)} \ldots = 2$,

which was first found in texts in many languages[6], namely, the most frequent word appears in long texts twice as many times as the second most frequent word, and the second most frequent word appears twice as many times as the fourth most frequent word and so forth.

Eq. (5) can be written also in a different way, namely,

$$\varphi(n) = \beta^{-1} \ln(1 + \frac{1}{n}) \qquad (6)$$

To obtain the normalized relative frequency of a box with $n$ balls, we divide the frequency by the total number of boxes with balls,

$\sum_{n=1}^{N} \varphi(n) = \beta^{-1} \sum_{n=1}^{N} \ln(1 + \frac{1}{n}) = \beta^{-1} \ln(N + 1)$

Which yields,

$$\phi(n, N) = \ln(1 + \frac{1}{n}) / \ln(N + 1) \qquad (7)$$

Choosing the value of $N$ reflects the difficulties arises from the subjective nature of the entropy. If we go back to our example of 12 balls distributed in 3 boxes, $N$ may be 3 or 12, depending on the question that we ask. If we want to know the distribution of particles in 3 boxes, than $N = 3$ and the equilibrium distribution of the balls between the boxes is 50%, 29%, 21%. However a box may have a probability to have any number of balls up to 12. In this case, when we substitute $N = 12$ in Eq.(7), we calculate the probability of finding a given possible number of balls. For example, the probability of finding a box with 5 balls is $\ln(1.2)/\ln(13)$, which is about 7.1%. The confusion between the these two kind of probabilities is the reason why most people will conclude that if million balls distributed randomly in 3 boxes the best bet is to find 333,333 balls in a box, while the probability to find exactly 333,333 balls in a box is only about $2.17 \times 10^{-5}$ %.

We call Eq.(7) Planck-Benford law[3]. It is worth noting that in Eq. (7) the Lagrange multipliers were cancelled out in the normalization. This makes this law universal, as it independent in the total number of balls and the total number of boxes. This is the reason that Benford's law and Zipf's law were discovered so many years ago. However, the Lagrange multiplier $\beta$ does not cancelled out in the density distribution function - given by

$$\rho(n, \beta) = n\varphi(n) = \beta^{-1} n \ln(1 + \frac{1}{n}) \qquad (8)$$

In Fig 1. the density distribution function $\rho = n\varphi(n)$ of 20 balls and $\beta = 1$ that was calculated from Eq. (8) is presented by the upper orange curve. As is seen, pretty fast, the plateau of Zipf's law in which $n\varphi(n) \approx 1$ appears. The lower black curve is the normalized relative frequency of the boxes having $n$ balls which were calculated from Eq. (7). The boxes with a smaller number of balls have higher frequencies.

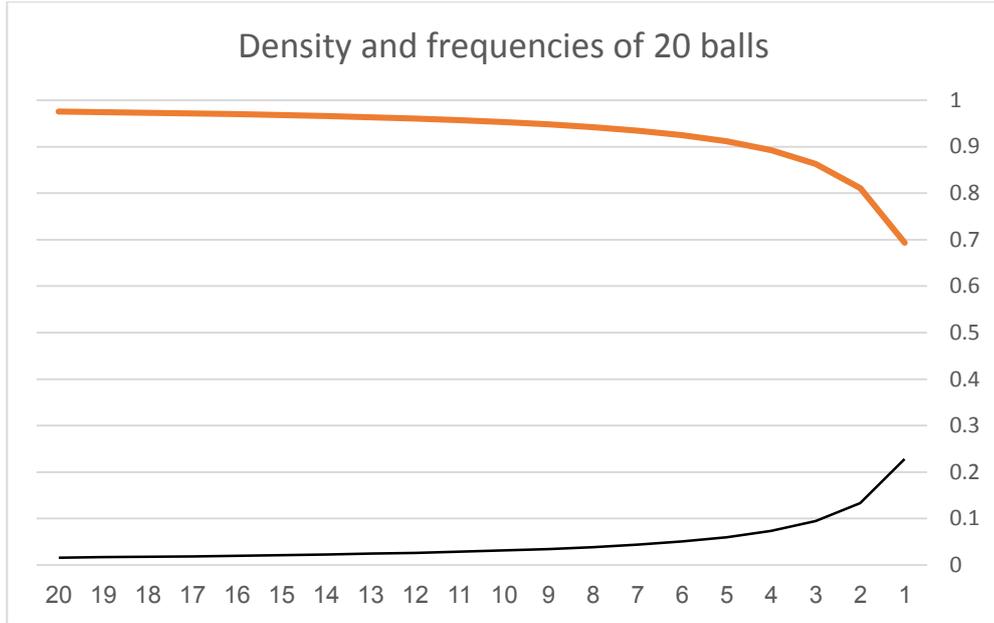

**Fig 1.** *The upper orange curve is the density distribution $\rho = n\varphi(n) = \beta^{-1} = 1$ of 20 balls. The lower black curve is the normalized relative probability of the boxes with $n$ balls [Eq.(7)] for $N = 20$.*

Zipf found his ratio of the frequencies of words in 1949 [6,7]. Zipf attributed the phenomenon to psychological argumentations of efficiency. It is shown here that the reason is probability.

### *Example 2* Applications of Planck-Benford law: Benford's Law

When we substitute $N = 9$ in Eq. (7) we obtain Benford's law.

$$\phi(n) = \ln(1 + \tfrac{1}{n})/\ln 10 = \log(1 + \tfrac{1}{n}) \tag{9}$$

Namely, Benford's law is the maximum entropy distribution when the boxes (digits) may have 1, 2, 3, 4, 5, 6, 7, 8 or 9 balls in a box. The Planck-Benford distribution excludes the empty boxes (the digits zero) as was explained previously and this is the reason why Benford's law is attributed erroneously to the first digit which is never 0. It is worth noting that Zipf's law and Benford's law are about normalized frequencies and are not density distributions. Usually we are interested in the density distribution function $\rho(n, \beta)$ which is the distribution of the balls, namely the fraction of the boxes having $n$ particles multiplied by $n$. The distribution function is plotted in the upper curve of Fig 1.

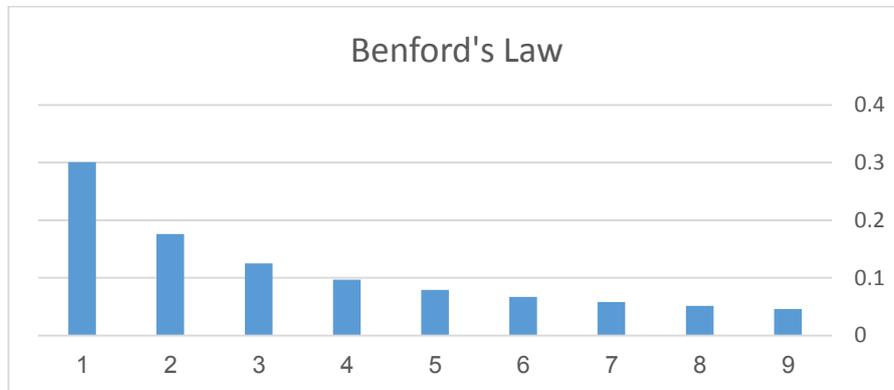

Fig. 2 *The normalized relative frequencies of boxes with $n$ balls where $1 \leq n \leq 9$.*

It should be noted that $N$ is the number of the distinguishable boxes that have different number of balls. The number of the boxes in the ensemble which is the total number of digits in the file is not relevant. Benford's law was found in the normalized relative frequencies of occurrences of the first digits of logarithmic tables by Newcomb who also guessed empirically Eq. (9) in 1881 [8]. Later it was discovered to be correct in many random decimal data by Benford in 1938[9].

*Example 3* **Applications of Planck-Benford law: Prior distributions of Polls**

Eq. (7) may be useful for predicting the prior distribution of polls. For example, in the Israeli parliament there were (eventually) 10 parties competing for 120 seats in the March 2015 election. We can use Eq. (6) to find the equilibrium distribution of seats among the ten parties. This problem is different from Benford's law. Here, the number of the different boxes is equal to the number of boxes. In Benford's law, we ask what the normalized frequency of boxes having $n$ balls. In polls we ask how the seats will be divided among the parties. Here we take $\phi(n)$ of Planck-Benford and divide the seats according to the relative normalized frequency of the parties. Therefore, the number of seats of the $n$'s party is calculated by $M(n) = \text{Round}[120 \frac{\ln(1+\frac{1}{n})}{\ln(11)}]$.

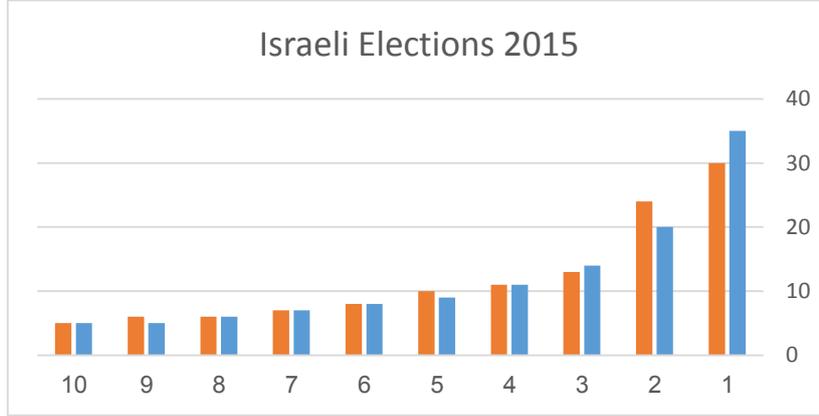

Fig. 3. *The distribution of 120 seats among 10 parties. The right blue bars are the Planck-Benford distribution and the left orange bars are the actual distribution of seats.*

Fig. 3 shows that the distribution of the seats in the Israel parliament obeys Planck Benford's law, which shows that no major forgery was made. It is worth noting that no free parameter was used in this calculation.

*Example 4* **Planck law and Planck-Einstein assumption**

We have derived the probabilistic Planck equation, Eq.(5), for pure logical entities that we call balls. Now let us assume that each ball has an amount of energy $h'$, and that all the balls are moving at the same velocity $c$, where the velocity has units of boxes per second. In this case the spatial frequency of the appearance of the balls $\varphi$ will be replaced by the temporal frequency $\nu = c\varphi$, and the energy of the system becomes $E = Lnh'$. We obtain from the number of balls $P = Nn\varphi(n)$ the energy $E = Nnh\nu$ where $h \equiv h'/c$. We rewrite the Lagrange of Eq.(4),

$$F(n) \cong L[(n+1)\ln(n+1) - n \ln n] + \beta'(E - Nnh\nu) + \alpha(L - N\nu/c) \qquad (10)$$

which yields the famous Planck equation,

$$n = \frac{1}{\exp(\beta h\nu(n)) - 1} \qquad (11)$$

The present probabilistic outcome of Planck-Einstein assumption that $E(n) = E/N = nh\nu$, which means that the higher the energy of a photon, the higher its frequency, is a direct outcome of the constraint on the number of balls of Eq.(3), namely, $P = Nn\varphi(n)$. If we view a single oscillator as the group of all the boxes with the same number of balls $n$, then $N$ is the number of the (longitudinal) modes of the radiation. In the Zipf's law regime all modes have the same energy however, this is not true for $n < 1$, in the quantum limit. In the quantum limit, the modes occupation number behave like in sparse systems and have exponential energy probability decay like in ideal gas.

When the modes have the same energy, smaller number of balls in a box means more energy per ball, and since boxes with a smaller number of balls appear more frequently, the frequency of the radiation is higher. It means that Planck's empirical assumption that higher the frequency higher the photon energy, that was empirically invoked by him to obtain the Blackbody Radiation formula, namely $E(n = 1) = h\nu$, is an outcome of Zipf's law. Planck's equation and his assumption were published by Planck in 1901 [11].

### *Example 5* balls with property: Bose-Einstein distribution

Now let us assume that the balls are static and have a simple property $\varepsilon$. Now the constraint on the number of balls and the constraint on the energy are simply $P = Ln$ and $E = \varepsilon Ln$. The Lagrange equation becomes,

$$F(n) \cong L[(n+1)\ln(n+1) - n\ln n] + \beta(E - \varepsilon Ln) + \alpha(P - Ln) \tag{12}$$

We ask that $\partial F(n)/\partial n = 0$.

and obtain

$\ln(n+1) - \ln n - \beta\varepsilon - \alpha = 0$, which yields

$n(\varepsilon, \alpha, \beta) = \frac{1}{\exp(\beta\varepsilon + \alpha) - 1}$

and the property distribution function $\rho(\varepsilon) = \varepsilon n(\varepsilon)$ is,

$$\rho(\varepsilon, \alpha, \beta) = \frac{\varepsilon}{\exp(\beta\varepsilon + \alpha) - 1} \tag{13}$$

Eq. (13) is the Bose-Einstein distribution. We saw that photons that are eventually pure energy behave like balls in boxes. However objects with a chemical potential in which the number of balls and their energies are independently constrained have chemical potential.

In Fig. 4 we provide three plots of Eq. (13), for $\rho(\varepsilon) = \varepsilon n(\beta = 1, \alpha = 0)$ which is a Planck function, and $\rho(\varepsilon) = \varepsilon n(\beta = 1, \alpha = 0.1)$ and $\rho(\varepsilon) = \varepsilon n(\beta = 1, \alpha = 1)$ which are Bose Einstein functions.

In the plots the functions are multiplied by an arbitrary number-10, namely the curves are $\rho(x) = \frac{10x}{\exp(x + \alpha) - 1}$ with $\alpha = 0$, which is the Planck equation, below we set $\alpha = 0.1$ (Bose Einstein), and the lowest is the blue with $\alpha = 1$.

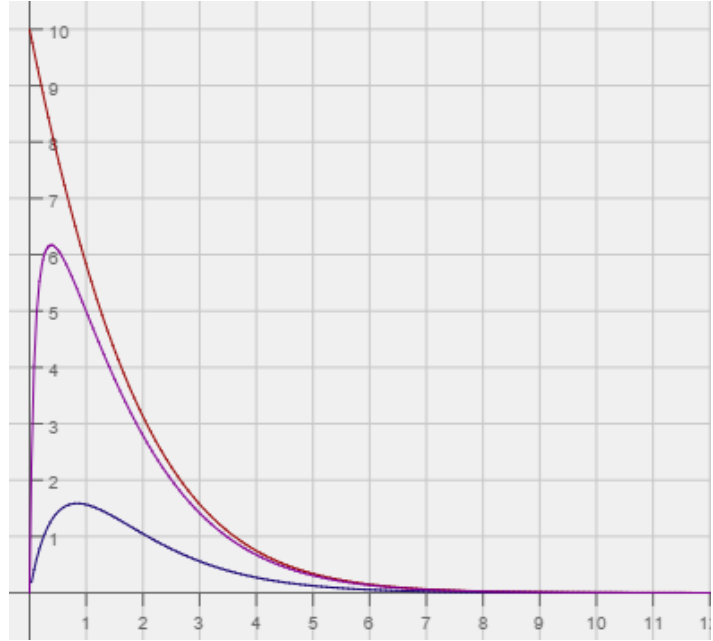

Fig. 4 *The upper red curve is the Planck distribution and the purple and blue below are the Bose-Einstein distributions with finite chemical potentials.*

The upper plot of Fig. 1, the upper plot of Fig 4. and the plot of Blackbody radiation that appears in the textbooks looks entirely different, these differences worth an explanation. In Fig. 1 we plotted $\rho(n,\beta) = n\varphi(n)$. Namely the density of balls versus the number of balls in a box. We see that the density of the balls at high $n$'s is independent of $n$. In Fig. 4 the energy density $\rho(\varepsilon,\beta) = \varepsilon\varphi(\varepsilon)$ is plotted versus the energy for high energies $\varepsilon > 1$ which also means $n \ll 1$. In this region the density decay exponentially, these energies are not shown in Fig. 1 at all. At very small photon energy where $n$ is large the density approach, according to Eq. (8), to $1/\beta$ as $\lim_{\varepsilon \to 0}\rho(\varepsilon,\beta) = 1/\beta$. It should be noted that at high energies of the balls in which $n \leq 1$ all the distributions in nature decay exponentially and the most of the balls have the average "property" value. However this is not true for $n \gg 1$ in the Zipf's law regime.

In most textbook the energy flux of the radiation is plotted and therefore the density function is multiplied by the number of the radiation modes per a given volume. The number of modes $N$ in a given volume is proportional to $\lambda^{-3}$, where $\lambda$ is the wavelength of the radiation. Therefore $N$ proportional to $\nu^3$ or $\varepsilon^3$. When we multiply the density $\rho(\varepsilon,\beta)$ by $\varepsilon^3$ namely $\varepsilon^4\varphi(\varepsilon)$ the bell like blackbody radiation curve that in common in the literature is obtained.

In the lower graph of Fig. (4) we see that the chemical potential has a dramatic effect that causes the Planck monotonically decreasing function to be similar to Maxwell-Boltzmann function.

Bose-Einstein distribution was suggested by Bose in 1924 [12] and by Einstein in 1925 [13].

*Sparse systems*

*Example 6* **Fermi-Dirac and Maxwell-Boltzmann distributions**

Now we look at a system in which no more than one ball can exist in a box. The number of the configurations is given by

$W = L!/(L-P)!P!$ and $S = \ln W \cong -L[n \ln n + (1-n)\ln(1-n)]$ with $n < 1$. Eq. (12) becomes,

$$F(n) \cong -L[n \ln n + (1-n)\ln(1-n)] + \beta\,(E - \varepsilon Ln) + \alpha(P - Ln) \qquad (14)$$

It should be noted that since $n < 1$, the sign of the entropy expression is negative. We ask that $\partial F(n)/\partial n = 0$.

We obtain $\ln\frac{1-n}{n} = \beta\varepsilon + \alpha$  or,

$n = \frac{1}{\exp(\beta\varepsilon+\alpha)+1}$ at the limit where $\beta\varepsilon$ is large, one obtains that $n \propto e^{-\beta\varepsilon}$ which is called the canonical distribution. The density distribution function is given by,

$$\rho(\varepsilon,\alpha,\beta) = \frac{\varepsilon}{\exp(\beta\varepsilon+\alpha)+1} \tag{15}$$

When $\alpha = 0$ and $\beta\varepsilon$ is high, Eq.(15) is called the Maxwell-Boltzmann distribution and when $\alpha \neq 0$ it is called Fermi-Dirac distribution. In the examples below we chose the values $\rho(x) = 10x\exp(-x)$, $\rho(x) = \frac{10x}{\exp(x)+1}$ and $\rho(x) = \frac{10x}{\exp(x+1)+1}$.

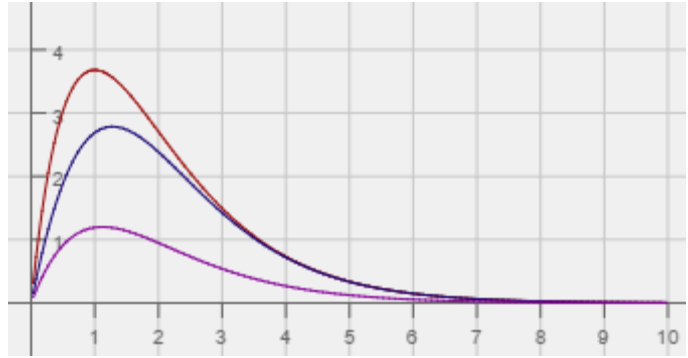

Fig. 5. *The upper red distribution is the Maxwell-Boltzmann, the blue distribution below is the Fermi-Dirac without chemical potential, and the purple distribution is the Fermi-Dirac with a chemical potential. All these distributions look similar.*

We see that the three curves are very similar and can easily be confused.

Maxwell-Boltzmann Distribution was first suggested by Maxwell in 1860 [19]. Fermi-Dirac distribution was suggested by Fermi and Dirac in 1926 [20, 21].

## Discussion

The formalism of maximizing the entropy using Lagrange multipliers technique is widely used in statistical mechanics to find thermal equilibrium energy distributions i.e. [4]. Ensemble of a fixed amount of particles, energy and volume is called a microcanonical ensemble. Here we strip the energy and volume off their physical identities and remain only with balls and boxes obeying the statistical second law, namely, equal probabilities to the boxes and equal probabilities to all the configurations of the balls in the boxes. An equal probability for any box yields an equal distribution of particles in the boxes. However, an equal distribution is not a maximum entropy solution. In nature, a statistical system tends to reach the equilibrium distribution in which entropy is maximum. To maximize the entropy one needs to add an equal probability requirement to the microstates.

Zipf's law was found by Zipf in the frequency of words in texts in many languages. Benford's law which is known also as the "first digit law" was found in the first digit of logarithmic tables (Benford's law is true for any digit in random decimal file regardless of its location). Many scientific papers were published about these distributions which are, in fact, merely different outcomes of the old Planck law. Benford's law is used for fraud detection in tax reports [18].

Similarly, it is possible to use Planck-Benford law described in this article for fraud detection in elections and surveys in addition to its predictive value. Moreover if we have different companies competing in the same market we can see if there is monopolistic power that distort the equilibrium distribution obtained by Planck-Benford law. If this theory would be applied in economics, each corporate would be able to calculate the equilibrium distribution of salaries among its employees. If the compensation distribution fits equilibrium, probably the employees will accept the salaries' gaps. If the salary distributions fits equilibrium, than $\boldsymbol{\beta}$ of the company's salaries distribution can be calculated and be compared to other companies as well as to $\boldsymbol{\beta}$ of the whole country salary distribution. In the present probability theory $\boldsymbol{\beta}$ replaces the reciprocal value of the average. In a recent paper [22] a network was described as a microcanonical ensemble. In a network, the boxes are the pairs of sites and the balls are the number of links in the pair. It was shown that high number of balls (links) in equilibrium is equivalent to high temperature ensemble. It was found that moving a ball-link from a hot net to a cold one increases entropy. However, moving a site which reduces the number of boxes from a hot net to a cold net decreases entropy [22,23]. These results explain the tendency of sites-people to emigrate from poor countries to the rich ones as is seen in the immigration waves from Africa to Europe and from Mexico to USA. Similarly this explains the tendency of links-money to flow from rich countries to poor ones as seems in the flow of money from USA to China and from Germany to Greece. These processes are an outcome of the present statistical theory.

A surprising result of this theory is that by adding a property to the balls we obtain a chemical potential. In physics, chemical potential is known as a change in the potential energy due to chemical reaction or phase transition. In fact chemical potential is an outcome of the constraint on the number of boxes. Increasing $\boldsymbol{\alpha}$ means increasing the number of boxes, or decreasing the average number of balls per box. Decreasing the average number of balls bring the system to be more similar to the canonical distribution of Maxwell-Boltzmann that we are so accustomed with. In Fig. (4) and Fig. (5) we see that in Bose-Einstein distribution the maximum of the functions is at $\boldsymbol{\alpha}$, which means that the average property of the balls are determined by the number of the boxes. In Fig. (5) it is seen that in sparse system most of the particles are found in the average which is determine by $\boldsymbol{\beta}$. When the property is energy, $\boldsymbol{\alpha}$ is potential energy. However, when the property is money, we have a similar "potential money" that may enable economists to treat financial processes and financial phase transitions with the proven arsenal of the physicists.

It should be mentioned that Planck made historic assumption [11] that the energy of a photon is quantized; namely the energy of a radiation mode is $\boldsymbol{E(n) = nh\nu}$. Here this equation was obtained by assuming that each ball carries an amount of energy $\boldsymbol{h'}$ and the balls are moving at the same velocity $\boldsymbol{c}$. This paper shows that for small number of ball in a box, which means more energy per ball, according to Zipf's law the boxes have higher frequency. This suggest a statistical explanation to Planck's assumption that the energy of a photon is proportional to its frequency.

It is important to note that in all the derivations presented in this paper we use Stirling formula. The Stirling formula yields inaccurate values for small numbers. In the example of 12 balls distributed randomly in 3 boxes that was presented previously, the exact combinatorial calculation shows that the empty box has the highest probability (14%), the second most probable is a box with one ball (13%), and the least expected is a box with 12 balls (1%). However, using Planck-Benford law we obtain 27% for box with 1 ball 3% for box with 12 balls and no value for empty boxes. The reason for this discrepancy is that we have not constrained the number of boxes in the entropy maximization process. The Planck-Benford law is independent of the number of boxes. If we use combinatorial calculation for 12 balls in 3 boxes, or in 4 boxes or in 100 boxes we will obtain different values for any number of boxes. Planck-Benford law is universal in this aspect, however we give up the knowledge of the probability of empty boxes and accurate values for small number of boxes and balls.

## Summary

We present a simple and effective way to calculate many real-life probabilities and distributions such as votes in election polls, market share distributions among competing companies, wealth distributions among populations and many others. By superseding classic probability theory with the maximum entropy principle we are able to calculate these complex distributions easily, and empirical real-life results strongly corroborate this novel approach. It is shown here that in addition this new approach yields the well-known distributions of Zipf's Law (with Pareto 80:20 law being its outcome), Benford's Law (which is the unequal distribution of digits in random data files), as well as energy distributions among particles in physics. This paper unifies everyday probability calculations done by surveyors, gamblers and economists with the calculations done by physicists, therefore enabling the applications of the methodology of statistical physics in economics and the life sciences.

## Acknowledgment

We want to thank H. Kafri and Y. Kafri for many comments. Special thanks to Alex E. Kossovsky for many comments and especially for the combinatorial calculations of balls and boxes.